\documentclass[prl,aps,twocolumn,superscriptaddress,showpacs,showkeys,amsmath,amssymb,floatfix]{revtex4}
\usepackage{booktabs,graphicx,verbatim}
\usepackage[colorlinks=true,citecolor=blue,filecolor=blue,linkcolor=blue,urlcolor=blue,pdftex]{hyperref}
\newcommand{\1}[1]{\, \mathrm{#1}} 
\newcommand{\n}[1]{\mathrm{#1}}    

\newcommand{\percent}{\%}

\newcommand{\degree}{{}^{\circ}}

\newcommand{\assergi}{\affiliation{INFN Laboratori Nazionali del Gran Sasso, Assergi, 67100, Italy}}
\newcommand{\columbia}{\affiliation{Physics Department, Columbia University, New York, NY 10027, USA}}
\newcommand{\coimbra}{\affiliation{Department of Physics, University of Coimbra, R. Larga, 3004-516, Coimbra, Portugal}}
\newcommand{\houston}{\affiliation{Department of Physics and Astronomy, Rice University, Houston, TX 77005 - 1892, USA}}

\newcommand{\losangeles}{\affiliation{Physics \& Astronomy Department, University of California, Los Angeles, USA}}
\newcommand{\munster}{\affiliation{Institut f\"ur Kernphysik, Westf\"alische Wilhelms-Universit\"at M\"unster, 48149 M\"unster, Germany}}
\newcommand{\shanghai}{\affiliation{Department of Physics, Shanghai Jiao Tong University, Shanghai, 200240, China}}
\newcommand{\subatech}{\affiliation{SUBATECH, Ecole des Mines de Nantes, Universit\'e de Nantes, CNRS/IN2P3, Nantes, France}}
\newcommand{\zurich}{\affiliation{Physics Institute, University of Z\"{u}rich, Winterthurerstr. 190, CH-8057, Switzerland}}

\begin{document}

\title{First Dark Matter Results from the XENON100 Experiment}

\author{E.~Aprile}\columbia
\author{K.~Arisaka}\losangeles
\author{F.~Arneodo}\assergi
\author{A.~Askin}\zurich
\author{L.~Baudis}\zurich
\author{A.~Behrens}\zurich
\author{E.~Brown}\losangeles
\author{J.~M.~R.~Cardoso}\coimbra
\author{B.~Choi}\columbia
\author{D.~B.~Cline}\losangeles
\author{S.~Fattori}\assergi
\author{A.~D.~Ferella}\zurich
\author{K.-L.~Giboni}\columbia
\author{K.~Bokeloh}\munster
\author{A.~Kish}\zurich
\author{C.~W.~Lam}\losangeles
\author{J.~Lamblin}\subatech
\author{R.~F.~Lang}\columbia
\author{K.~E.~Lim}\columbia
\author{J.~A.~M.~Lopes}\coimbra
\author{T.~Marrod\'an~Undagoitia}\zurich
\author{Y.~Mei}\houston
\author{A.~J.~Melgarejo Fernandez}\columbia
\author{K.~Ni}\shanghai
\author{U.~Oberlack}\houston
\author{S.~E.~A.~Orrigo}\coimbra
\author{E.~Pantic}\losangeles
\author{G.~Plante}\email[]{guillaume.plante@astro.columbia.edu}\columbia
\author{A.~C.~C.~Ribeiro}\coimbra
\author{R.~Santorelli}\zurich
\author{J.~M.~F.~dos Santos}\coimbra
\author{M.~Schumann}\zurich\houston
\author{P.~Shagin}\houston
\author{A.~Teymourian}\losangeles
\author{D.~Thers}\subatech
\author{E.~Tziaferi}\zurich
\author{H.~Wang}\losangeles
\author{C.~Weinheimer}\munster

\collaboration{XENON100 Collaboration}\noaffiliation

\begin{abstract}
The XENON100 experiment, in operation at the Laboratori Nazionali del Gran Sasso in Italy, is designed to search for dark matter WIMPs scattering off $62\1{kg}$ of liquid xenon in an ultra-low background dual-phase time projection chamber. In this letter, we present first dark matter results from the analysis of 11.17 live days of non-blind data, acquired in October and November 2009. In the selected fiducial target of $40\1{kg}$, and within the pre-defined signal region, we observe no events and hence exclude spin-independent WIMP-nucleon elastic scattering cross-sections above $3.4\times10^{-44}\1{cm}^2$ for $55\1{GeV/c^2}$ WIMPs at $90\percent$ confidence level. Below $20\1{GeV/c^2}$, this result constrains the interpretation of the CoGeNT and DAMA signals as being due to spin-independent, elastic, light mass WIMP interactions.
\end{abstract}



\maketitle

A vast array of observational evidence suggests that $83\percent$ of the matter in the universe is in some unknown form called dark matter~\cite{Komatsu:2008hk}. Extensions of the Standard Model of particle physics that aim at addressing some of the puzzles associated with the electroweak scale predict stable Weakly Interacting Massive Particles (WIMPs), that can be thermally produced in the early universe and become ideal dark matter candidates~\cite{Bertone2005279}. One method to detect WIMPs is to measure the energy they deposit in a detector by scattering off target nuclei.

XENON100 is a new ultra-low background detector developed to continue the XENON dark matter search program with liquid xenon (LXe) as WIMP target and detection medium. Like XENON10~\cite{Angle:2007uj}, it is a three-dimensional (3D) position-sensitive dual-phase (liquid/gas) time projection chamber (TPC) filled with ultra-pure LXe.  Particle interactions in the sensitive LXe volume are measured with two arrays of photomultiplier tubes (PMTs), which simultaneously detect the primary scintillation (S1) and the ionization signal (S2), via the proportional scintillation mechanism~\cite{Lansiart197647}. The 3D event localization allows ``fiducialization'' of the TPC, keeping only an inner volume in which the background rate is drastically reduced. The different ionization density of nuclear recoils, from neutrons or WIMPs, and electronic recoils, from $\gamma$ or $\beta$ backgrounds, leads to a different S2/S1 ratio, which can be used to discriminate the two types of recoils.

The detector uses 161~kg of LXe, divided in two concentric cylindrical volumes. These are physically and optically separated from each other by polytetrafluoroethylene (PTFE) panels on the side, a PMT array on the bottom, and a stainless steel diving bell on the top. The bell allows the inner liquid level to be set independently of the outer one. The PTFE panels define the TPC with $30.5\1{cm}$ diameter and $30.6\1{cm}$ height, acting also as an efficient UV reflector. Four stainless steel meshes provide the electric field to drift ionization electrons in the liquid, extract them to the gas phase, and accelerate them to produce proportional scintillation. A drift field of $530\1{V/cm}$ has been used for the measurements reported here.

The PMTs are $2.5\1{cm} \times 2.5\1{cm}$ metal-channel type (R8520-AL) specifically developed in
collaboration with Hamamatsu\:Co.\ for operation in LXe, with a quantum efficiency of about $30\percent$ at 178~nm and with low intrinsic radioactivity (\textless~1 mBq/PMT in $^{238}\n{U}$/$^{232}\n{Th}$). The 80~PMT array at the bottom of the TPC is immersed in the liquid to efficiently detect the S1 signal, while another array of 98~PMTs is placed in the xenon gas above the anode so that the hit pattern of an S2 signal can be used to reconstruct the event position in ($x,y$). The interaction depth ($z$) in the detector can be computed from the time difference between S1 and S2 pulses with resolution \textless~$2\1{mm}$. The outermost LXe volume is used as an active veto, instrumented with 64~PMTs. The energy threshold of the veto has been measured to be better than $200\1{keV_{ee}}$ (keV electron-equivalent~\cite{Aprile:2008rc} as inferred from $^{137}\n{Cs}$). The signals from all~242 PMTs are digitized at $100\1{MS/s}$ and $40\1{MHz}$ bandwidth. The trigger is provided by the summed signal of 84~central PMTs, low-pass filtered with $1\1{MHz}$. Given the strong amplification in the gas proportional region, at low energies the trigger is given by the S2 pulse, with an efficiency $>99\percent$ above 300~photoelectrons (PE).

The detector has been deployed underground at the Laboratori Nazionali del Gran Sasso (LNGS), where the muon flux is reduced by a factor $10^6$, thanks to the average 3600 m water equivalent of rock overburden. The LXe is contained in a double walled, vacuum insulated, stainless steel cryostat. A $200\1{W}$ pulse tube refrigerator (PTR) continuously liquifies the gas circulated through a hot getter and maintains the liquid at $-91\degree \n{C}$. The PTR system is installed outside a passive shield to achieve a lower radioactive background in the target. This shield consists of a $20\1{cm}$ thick layer of lead and a $20\1{cm}$ thick layer of polyethylene within, to attenuate the background from external $\gamma$-rays and neutrons, respectively. The shield structure rests on a $25\1{cm}$ thick slab of polyethylene and is surrounded on the top and three sides by a $20\1{cm}$ thick water layer to lower the background contribution from neutrons from the cavern rock. A $5\1{cm}$ thick layer of copper covers the inner surface of the polyethylene to attenuate the gamma background due to its radioactivity. Calibration sources ($^{57}\n{Co}$, $^{60}\n{Co}$, $^{137}\n{Cs}$, $^{241}\n{AmBe}$) are inserted through a copper tube which penetrates the shield and circles around the detector in the middle of the drift region.

The gas used in XENON100 has been processed through a distillation column to reduce the $^{85}$Kr background to $33\1{\mu Bq/kg}$, as measured with delayed $\beta$-$\gamma$ coincidences~\cite{Aprile:2010bt}. Assuming an isotopic abundance of $10^{-11}$, this $^{85}\n{Kr}$ contamination corresponds to $143^{+130}_{-90}\1{ppt}$ (mol/mol), at $90\percent$ confidence, of natural Kr.

PMT gains are measured under single PE conditions using light emitting diodes (LEDs) coupled to optical fibers which illuminate the TPC and veto volumes. The PMT gains, equalized to $1.9\times10^6$ at the beginning of the run, are regularly monitored and are stable within $\pm2\percent$ ($\sigma/\mu$).

Event positions are calculated using three independent algorithms, based on $\chi^2$ minimization, Support Vector Machine (SVM) regression, and a Neural Network (NN). We take the PMT gains into account and correct for non-uniformities of the drift field as inferred from a finite element simulation. The three algorithms give consistent results for radii $r < 14\1{cm}$ with an ($x,y$) resolution better than $3\1{mm}$, as measured with a collimated $\gamma$ source. This motivated the choice, for the present analysis, of a $40\1{kg}$ fiducial volume as a cylinder of radius $13.5\1{cm}$ and height $24.3\1{cm}$.

Corrections for the spatial dependence of the S1 light collection in the TPC are obtained by irradiating the detector at different azimuthal positions with an external $^{137}\n{Cs}$ source and computing the average light yield in $1\1{cm}\times2.5\1{cm}$ ($r,z$) cells. The average light yield of the whole TPC for $^{137}\n{Cs}$ $662\1{keV_{ee}}$ $\gamma$-rays is $1.57\1{PE/keV_{ee}}$ at a field of $530\1{V/cm}$. The spatial correction is also inferred using $40\1{keV_{ee}}$ $\gamma$-rays produced via the inelastic reaction $^{129}\n{Xe}\left({n, n'\gamma}\right)^{129}\n{Xe}$, together with $80\1{keV_{ee}}$ $\gamma$-rays from $^{131}\n{Xe}\left({n, n'\gamma}\right)^{131}\n{Xe}$, during the calibration of the detector with an external $^{241}\n{AmBe}$ source. These $\gamma$-rays are more uniformly distributed in the sensitive volume due to the larger neutron mean free path. In addition, $164\1{keV_{ee}}$ and $236\1{keV_{ee}}$ $\gamma$-rays are produced following the same neutron calibration from the decay of metastable $^{131m}\n{Xe}$ and $^{129\text{m}}\text{Xe}$, respectively. The $164\1{keV_{ee}}$ line is also used to infer the S1 spatial dependence. The corrections inferred from these independent calibrations differ by less than $3\percent$ and improve the energy resolution ($\sigma/E$) at $662\1{keV_{ee}}$ from $24\percent$ to $13\percent$ using the scintillation signal alone.

Calibrations with $^{137}\n{Cs}$ were taken daily during the data taking presented here, to infer the electron lifetime and to subsequently correct the S2 signal for its drift time dependence. The electron lifetime increased from $154\1{\mu s}$ to $192\1{\mu s}$, resulting in the average S2 $z$-correction decreasing from $75\percent$ to $60\percent$. The S2 signal is also corrected for its $(x,y)$ variation, mostly due to light collection effects near the edge of the TPC. This dependence is determined using the $40\1{keV_{ee}}$ $\gamma$-rays from the neutron calibration data and computing the proportional scintillation light yield in $2\1{cm}\times2\1{cm}$ $(x,y)$ cells. Only insignificant differences ($<2\percent$) were observed between corrections obtained using other calibration datasets of various $\gamma$-ray energies ($164\1{keV_{ee}}$, $662\1{keV_{ee}}$). The energy resolution ($\sigma/E$) at $662\1{keV_{ee}}$ using the S2 signal alone is improved from $7.3\percent$ to $6.5\percent$ after applying the S2 spatial corrections.

\begin{figure}[!htb]
\begin{center}\includegraphics[width=1\columnwidth]{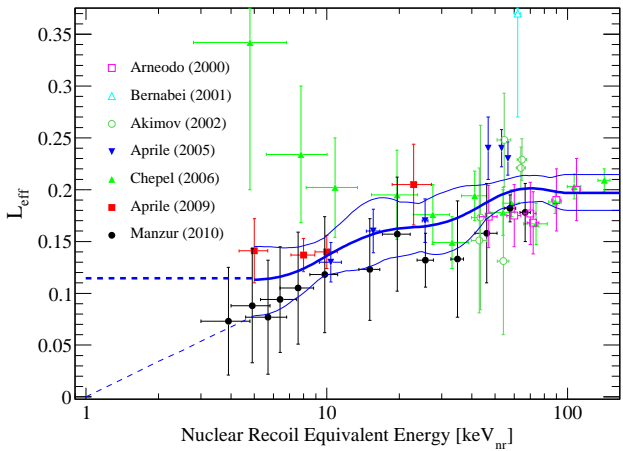}
\caption{Global fit to all $\mathcal{L}_{\text{eff}}$ measurements with fixed neutron energies between $5\1{keV_{nr}}$ and $100\1{keV_{nr}}$, together with $90\percent$ confidence contours (solid lines) and extrapolations to lower energies (dashed lines).}\label{fig:leff}
\end{center}\end{figure}

The nuclear-recoil equivalent energy $E_{\n{nr}}$ in LXe is conventionally computed from the scintillation signal, $S1$, using $E_{\n{nr}} = S1/L_y \cdot 1/\mathcal{L}_{\text{eff}} \cdot S_{\n{ee}}/S_{\n{nr}}$, where $\mathcal{L}_{\text{eff}}$ is the scintillation efficiency of nuclear recoils relative to that of $122\1{keV_{ee}}$ $\gamma$-rays at zero field, and $S_{\n{ee}}$ and $S_{\n{nr}}$ are the electric field scintillation quenching factors for electronic and nuclear recoils, respectively, with measured values of 0.58 and 0.95~\cite{Aprile:2006kx}.  Since $122\1{keV_{ee}}$ $\gamma$-rays cannot penetrate far in the sensitive volume, their light yield $L_y$ at $530\1{V/cm}$ is calculated from a fit to all $\gamma$-ray lines mentioned above, yielding $L_y(122\1{keV_{ee}})=(2.20\pm0.09)\1{PE/keV_{ee}}$.  $\mathcal{L}_{\text{eff}}$ data measured at fixed neutron energies~\cite{Aprile:2008rc,Manzur:2009hp,Arneodo:2000vc}, shown in Fig.~\ref{fig:leff}, have less systematic uncertainty than those inferred from a comparison of neutron calibration spectra with Monte Carlo simulations.  Therefore, the energy dependence of $\mathcal{L}_{\text{eff}}$ and its uncertainty is determined here through a global cubic-spline fit to all data shown in Fig.~\ref{fig:leff} in the energy range with at least two measurements ($5-100\1{keV_{nr}}$). The spline knots are fixed at 5, 10, 25, 50 and $100\1{keV_{nr}}$.  Below $5\1{keV_{nr}}$, a constant extrapolation of the global fit, consistent with the trend reported in Aprile \textit{et al.}~\cite{Aprile:2008rc} and Sorensen \textit{et al.}~\cite{Sorensen:2008ec}, is used in this analysis. A logarithmic extrapolation of the lower $90\percent$ confidence contour to zero scintillation near $1\1{keV_{nr}}$, following the trend in Manzur \textit{et al.}~\cite{Manzur:2009hp}, is also shown in Fig.~\ref{fig:leff}.

\begin{figure}[!htb]
\begin{center}\includegraphics[width=1\columnwidth]{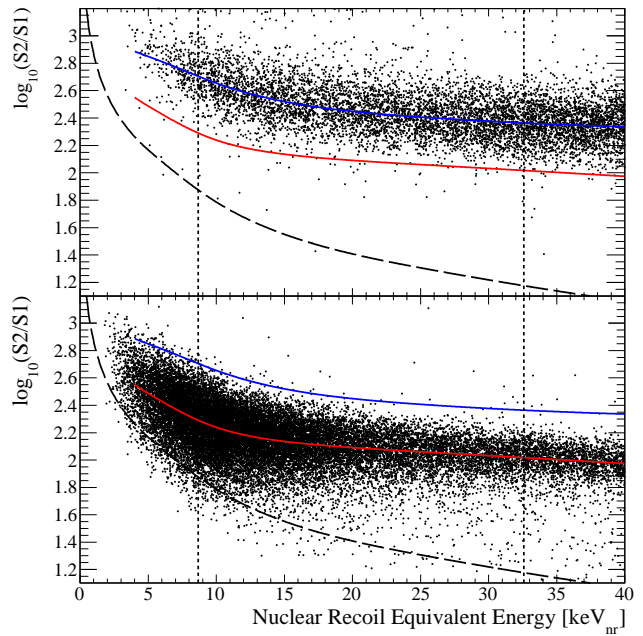}
\caption{Electronic (top) and nuclear (bottom) recoil bands from $^{60}\text{Co}$ and $^{241}\n{AmBe}$ calibration data, respectively, after data selection and the $40\1{kg}$ fiducial volume cut. Colored lines correspond to the median $\log_{10}(\n{S2/S1})$ values of the electronic (blue) and nuclear (red) recoil bands. The WIMP search energy window $8.7-32.6\1{keV_{nr}}$ (vertical, dashed) and S2 software threshold of $300\1{PE}$ (long dashed) are shown.}\label{fig:bands}
\end{center}\end{figure}

Data selection criteria are motivated by the physical properties of xenon scintillation light, the characteristics of proportional light signals, and the expected WIMP-induced single-scatter nuclear-recoil signature. Cuts were developed and tested on calibration data, specifically on low energy electronic recoils from Compton scattered $^{60}\n{Co}$ $\gamma$-rays and nuclear recoils from $^{241}\n{AmBe}$. In particular, a two-fold PMT coincidence is required in a $20\1{ns}$ window for the S1 signal and events which contain more than a single S1-like pulse are discarded. This allows true low energy events to be distinguished from events with random single photoelectrons from PMTs or accidental coincidences. For the S2 signal, a lower threshold of $300\1{PE}$ is set, corresponding to about 15~ionization electrons, and events are required to contain only one S2 pulse above this threshold. This rejects events with multiple scatters at different $z$ positions. In addition, the width of the S2 pulse is required to be consistent with what is expected from the inferred drift time due to diffusion of the electron cloud~\cite{doke1982}. Events that deposit energy in the veto volume in coincidence with the S1 signal in the TPC are also discarded. The regions of the digitized waveform away from S1 or S2 pulses are required to be free of extraneous PMT signals or noise. Finally, events outside the pre-defined fiducial volume are rejected.

\begin{figure}[!htb]
\begin{center}\includegraphics[width=1\columnwidth]{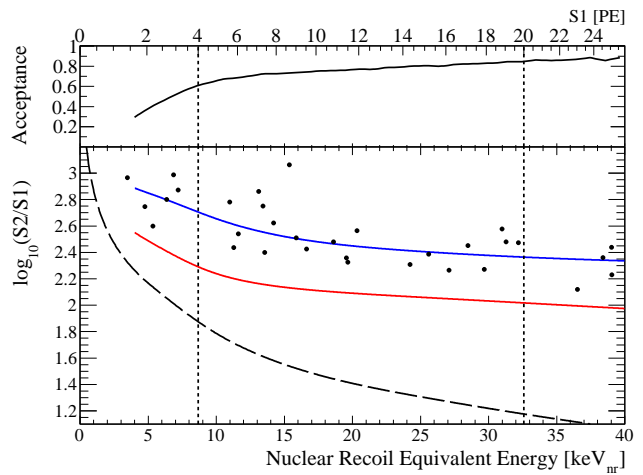}
\caption{Cut acceptance (top, not including 50\% acceptance from $\n{S2/S1}$ discrimination) and $\log_{10}(\n{S2/S1})$ (bottom) as functions of nuclear recoil energy for events observed in the $40\1{kg}$ fiducial volume during 11.17 live days. Lines as in figure~\ref{fig:bands}.}\label{fig:discrimination}
\end{center}\end{figure}

Background rejection in XENON100 is achieved through a combination of volume fiducialization and the identification of recoil species based on the ratio S2/S1 for electronic and nuclear recoils. Accurate knowledge of the response to both types of recoils is essential to define the signal region, to determine the signal acceptance, and to predict the expected leakage into the signal region. Statistics for the low energy electronic recoil calibration are accumulated at regular intervals with a $1\1{kBq}$ $^{60}\text{Co}$ source.  The response of XENON100 to elastic nuclear recoils was obtained by irradiating the detector with a $220\1{n/s}$ $^{241}\n{AmBe}$ source for $72\1{h}$. Fig.~\ref{fig:bands} shows the $\log_{10}(\n{S2/S1})$ distribution of single scatter electronic and nuclear recoils in the $40\1{kg}$ fiducial volume, as function of nuclear recoil energy. The energy window for the WIMP search is chosen between $8.7-32.6\1{keV_{nr}}$ ($4-20\1{PE}$). The upper bound is taken to correspond approximately to the one used for the XENON10 blind analysis~\cite{Angle:2007uj}, after recomputing the corresponding nuclear-recoil equivalent energy using the new $\mathcal{L}_{\text{eff}}$ parametrization from the global fit, shown in Fig.~\ref{fig:leff}. The lower bound is motivated by the fact that the acceptance of the S1 two-fold coincidence requirement is $>90\percent$ above $4\1{PE}$. The $\log_{10}(\n{S2/S1})$ upper and lower bounds of the signal region are respectively chosen as the median of the nuclear recoil band and the 300~PE S2 software threshold.

\begin{figure}[!htb]
\begin{center}\includegraphics[width=1\columnwidth]{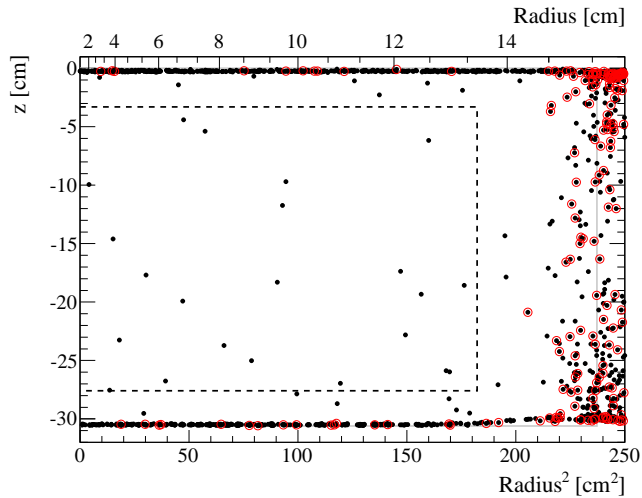}
\caption{Distribution of all events (dots) and events below the nuclear recoil median (red circles) in the TPC (grey line) observed in the $8.7-32.6\1{keV_{nr}}$ energy range during 11.17~live days. No events below the nuclear recoil median are observed within the $40\1{kg}$ fiducial volume (dashed).}\label{fig:rz}
\end{center}\end{figure}

A first dark matter analysis has been carried out, using 11.17~live days of background data, taken from October 20th to November 12th 2009, prior to the neutron calibration. Although this was not formally a blind analysis, all the event selection criteria were optimized based on calibration data only. The cumulative software cut acceptance for single scatter nuclear recoils is conservatively estimated to vary between $60\percent$ (at $8.7\1{keV_{nr}}$) and $85\percent$ (at $32.6\1{keV_{nr}}$) by considering all single-scatter events in the fiducial volume that are removed by only a single cut to be valid events (Fig.~\ref{fig:discrimination}). Visual inspection of hundreds of events confirmed that this is indeed a conservative estimate. Within the $8.7-32.6\1{keV_{nr}}$ energy window, 22 events are observed, but none in the pre-defined signal acceptance region (Fig.~\ref{fig:discrimination}). At $50\percent$ nuclear recoil acceptance, the electronic recoil discrimination based on $\log_{10}(\n{S2/S1})$ is above $99\percent$, predicting \textless~0.2 background events in the WIMP region. The observed rate, spectrum, and spatial distribution (Fig.~\ref{fig:rz}) agree well with a GEANT4 Monte Carlo simulation of the entire detector.

\begin{figure}[!htb]
\begin{center}\includegraphics[width=1\columnwidth]{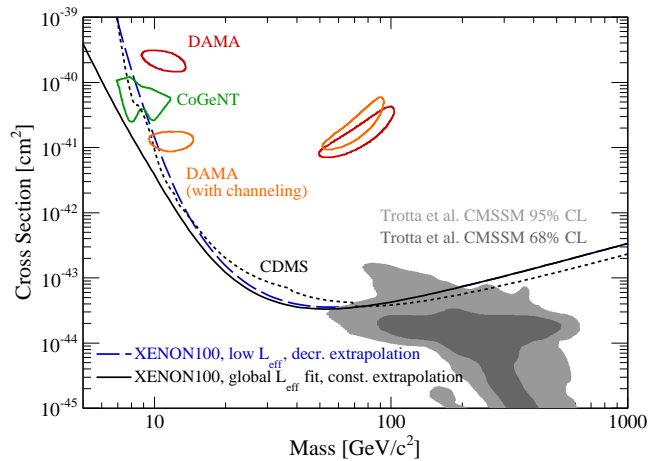}
\caption{$90\percent$ confidence limit on the spin-independent elastic WIMP-nucleon cross section (solid and long dashed), together with the best limit to date from CDMS~(dotted)~\cite{Ahmed:2010}, re-calculated assuming an escape velocity of $544\1{km/s}$ and $v_0 = 220\1{km/s}$. Expectations from a theoretical model~\cite{Trotta:2008bp}, and the areas ($90\percent$ CL) favored by CoGeNT (green)~\cite{Aalseth:2010vx} and DAMA (red/orange)~\cite{Savage:2008er} are also shown.}\label{fig:limit}
\end{center}\end{figure}

An upper limit on the spin-independent WIMP-nucleon elastic scattering cross section is derived based on the assumption of an isothermal WIMP halo with $v_0 = 220\1{km/s}$, density $0.3\1{GeV/c^2}$, and escape velocity $544\1{km/s}$~\cite{Lewin:1995rx}. We take a S1 resolution dominated by Poisson fluctuations into account and use the global fit $\mathcal{L}_{\text{eff}}$ with constant extrapolation below $5\1{keV_{nr}}$.  The acceptance-corrected exposure in the energy range considered, weighted by the spectrum of a $100\1{GeV/c^2}$ WIMP, is $172\1{kg} \cdot \n{days}$. Fig.~\ref{fig:limit} shows the resulting $90\percent$ confidence upper limit, with a minimum at a cross section of $3.4\times10^{-44}\1{cm}^2$ for a WIMP mass of $55\1{GeV/c^2}$. The impact of assuming the lower $90\percent$ confidence $\mathcal{L}_{\text{eff}}$ contour together with the extrapolation to zero around $1\1{keV_{nr}}$ is also shown. Our limit constrains the interpretation of the CoGeNT~\cite{Aalseth:2010vx} and DAMA~\cite{Savage:2008er} signals as being due to light mass WIMPs. These initial results, based on only 11.17 live days of data, demonstrate the potential of the XENON100 low-background experiment to discover WIMP dark matter.

We gratefully acknowledge support from NSF, DOE, SNF, the Volkswagen Foundation, FCT, and STCSM. We are grateful to the LNGS for hosting and supporting the XENON program. We acknowledge the contributions of T.~Bruch (UZH), K.~Lung (UCLA), A.~Manalaysay (UZH), and M.~Yamashita (U. Tokyo).


\end{document}